\documentclass{emulateapj}
\usepackage{graphicx}
\usepackage[usenames,dvips]{color}
\usepackage{courier}
\usepackage[T1]{fontenc}
\def\arcsec{$\,^{\prime\prime}$~}

\newcommand{\be}{\begin{equation}}
\newcommand{\bel}[1]{\begin{equation}\label{eq:#1}}
\newcommand{\ee}{\end{equation}}
\newcommand{\bd}{\begin{displaymath}} 
\newcommand{\ed}{\end{displaymath}}   
\newcommand{\bea}{\begin{eqnarray}}
\newcommand{\beal}[1]{\begin{eqnarray}\label{eq:#1}}
\newcommand{\eea}{\end{eqnarray}}

\newcommand{\eqref}[1]{\ref{eq:#1}}

\newcommand{\lsim }{{\lower0.8ex\hbox{$\buildrel <\over\sim$}}}
\newcommand{\gsim }{{\lower0.8ex\hbox{$\buildrel >\over\sim$}}}

\def\Chandra{${\it Chandra}$}

\def\simge{\mathrel{%
   \rlap{\raise 0.511ex \hbox{$>$}}{\lower 0.511ex \hbox{$\sim$}}}}
\def\simle{\mathrel{
   \rlap{\raise 0.511ex \hbox{$<$}}{\lower 0.511ex \hbox{$\sim$}}}}

\newcommand{\Msun}{\ifmmode {M_{\odot}}\else${M_{\odot}}$\fi}
\newcommand{\Lsun}{\ifmmode {L_{\odot}}\else${L_{\odot}}$\fi}
\newcommand{\Rsun}{\ifmmode {R_{\odot}}\else${R_{\odot}}$\fi}

\shorttitle{2.15 Hour Period for LMXB in Globular Cluster NGC 6652}
\shortauthors{Engel et al.}

\begin{document}
\title{A 2.15 Hour Orbital Period for the Low Mass X-Ray Binary XB 1832-330 in the Globular Cluster NGC 6652}  

\author{M.~C. Engel\altaffilmark{1}, C.~O. Heinke\altaffilmark{1,2}, G.~R. Sivakoff\altaffilmark{1}, K. G. Elshamouty\altaffilmark{1}, P.~D. Edmonds\altaffilmark{3}}

\altaffiltext{1}{Physics Dept., 4-183 CCIS, Univ. of Alberta, Edmonton AB T6G 2E1, Canada; mcengel@ualberta.ca}

\altaffiltext{2}{Ingenuity New Faculty; heinke@ualberta.ca}

\altaffiltext{3}{Harvard-Smithsonian Center for Astrophysics, 60 Garden Street, Cambridge, MA 02138, USA}


\begin{abstract}
We present a candidate orbital period for the low mass X-ray binary XB 1832-330 in the globular cluster NGC 6652 using a 6.5 hour Gemini South observation of the optical counterpart of the system. Light curves in g' and r' for two LMXBs in the cluster, sources A and B in previous literature, were extracted and analyzed for periodicity using the ISIS image subtraction package. A clear sinusoidal modulation is evident in both of A's curves, of amplitude $\sim$0.11 magnitudes in g' and $\sim$0.065 magnitudes in r', while B's curves exhibit rapid flickering, of amplitude $\sim$1 magnitude in g' and $\sim$0.5 magnitudes in r'. A Lomb-Scargle test revealed a 2.15 hour periodic variation in the magnitude of A with a false alarm probability less than 10$^{-11}$, and no significant periodicity in the light curve for B. Though it is possible saturated stars in the vicinity of our sources partially contaminated our signal, the identification of A's binary period is nonetheless robust.
\end{abstract}


\maketitle

\section{Introduction}\label{s:intro}
Globular clusters are efficient factories for the production of low-mass X-ray binaries (LMXBs), due to their high central densities and increased rates of stellar interactions \citep{Clark75,Verbunt06}. The dynamical formation mechanisms in globular clusters produce LMXBs with different characteristics than those in the rest of the galaxy, particularly in their overproduction of ultracompact (P$<$80 minutes) LMXBs requiring a degenerate donor star \citep{Deutsch00}. 
Numerous LMXBs have been detected in globular clusters around other galaxies \citep[e.g.][]{Sarazin01,Angelini01}, allowing identification of a metallicity dependence in their formation \citep{Kundu02}, better statistics of their dependence on cluster structural parameters \citep{Jordan07}, and high-quality X-ray luminosity functions \citep{Kim09,Zhang11}.  However, local globular cluster LMXBs are generally the only objects we can study in detail, obtaining key parameters such as binary orbital periods for use in population synthesis calculations \citep[e.g.][]{Fragos08,Ivanova08}.
As of 2011, 15 bright ($L_{X,peak}>10^{36}$ ergs/s) LMXBs in galactic globular clusters are known; of eleven known periods, five are ultracompact,  while two require ($P_{orb}>$15 hours) evolved donor stars   \citep{Verbunt06,Dieball05,Zurek09,Altamirano08,Altamirano10,Strohmayer10}. Identification of these orbital periods has generally required Hubble Space Telescope (HST) imaging, or X-ray studies, with the exception of the optically bright AC 211 in M15 \citep{Ilovaisky93}.

X-ray emission from the globular cluster NGC 6652 was detected by HEAO-1 in 1977-78 \citep{Hertz85}, at $L_X$(2-10 keV) $\sim10^{35}$ ergs/s, and then at $L_X\sim10^{36}$ ergs/s by ROSAT in 1990 \citep{Predehl91}. This LMXB has since been observed by BeppoSAX \citep{intZand98}, ASCA \citep{Mukai00}, and XMM \citep{Sidoli08} at $L_X\sim10^{36}$ ergs/s.  Since 1999 it has been monitored by RXTE's PCA instrument during the bulge scan program \citep{Swank01}\footnote{http://lheawww.gsfc.nasa.gov/users/craigm/galscan/main.html}, finding it roughly constant ($L_X\sim10^{36}$ ergs/s) until 2011, during which it declined by a factor of $\sim$3.

\citet{Deutsch98a} identified a candidate UV-bright, variable optical counterpart (their star 49) to the NGC 6652 LMXB at 2.3$\sigma$ from the ROSAT position.  \citet{Deutsch00} found a candidate 43.6 minute period in three orbits of HST $V$ and $I$ imaging, with a Fourier peak at 99.5\% confidence. However, only two HST orbits (each including $\sim$45 minutes of on-source time) are consistent with this period, while the third orbit shows strong flickering (see their Fig. 1), and \citet{Heinke01} argued the period was not convincing.  \citet{Deutsch00} note that the optical faintness of star 49 indicates a small accretion disk, and thus also suggests an ultracompact nature.

\Chandra\ high-resolution X-ray imaging of NGC 6652 has revealed several X-ray sources in the cluster, hereafter A through G in order of brightness \citep{Heinke01,Coomber11}. The bright LMXB A (XB 1832-330) was identified by \citet{Heinke01} with a different blue, variable star, showing tentative evidence for a sinusoidal period of either 0.92 or 2.22 hours. Deutsch's star 49 is the optical counterpart of B, a lower-luminosity ($L_X\sim4\times10^{33}$ ergs/s) X-ray source. A 5 kilosecond  \Chandra\  exposure revealed rapid flaring from B on timescales down to 100 seconds, from $L_X<2\times10^{33}$ ergs/s up to $L_X\sim10^{35}$ ergs/s.  B's long-term $L_X$ has appeared relatively constant, as it has been detected with ROSAT in 1994, and \Chandra\ in 2000, 2008, and 2011 at similar $L_X$s \citep{Coomber11,Stacey11b}.

A and B are somewhat unusual among globular cluster LMXBs in being located well outside the core of their globular cluster \citep{Verbunt06}, in less crowded regions potentially observable from the ground. This motivated us to search for optical periodicities in both A and B using Gemini-South.

\section{Data Reduction}\label{s:obs}

We observed NGC 6652 on 2011 May 2 for 6.5 hours with Gemini's GMOS-S CCD detector. Each exposure was 75 s in duration, and 172 images were taken alternately in the g' and r' filters.
Raw CCD data were prepared by the Gemini IRAF\footnote{http://iraf.noao.edu/} package task GPREPARE, and flat-field and bias corrections, as well as gain multiplication, were performed by the task GIREDUCE using Gemini-supplied calibration information\footnote{http://www.gemini.edu/sciops/instruments/gmos/}. The GMOS-S instrument contains three CCD chips, but only the image data from the central chip, containing the majority of the cluster, was used.

Photometry was performed on the images using the image subtraction package ISIS\footnote{http://www2.iap.fr/users/alard/package.html}, developed by Alard \& Lupton \citep{Alard98,Alard00} to deal with very crowded fields exhibiting spatially varying seeing and background levels. ISIS works by transforming images to a common seeing, subtracting these from one another, and performing photometry on the more sparsely populated subtracted images. To minimize PSF residuals in the subtracted image, ISIS uses a linear least squares fitting method to optimize a solution for a spatially-varying kernel which can be convolved with a reference image to recreate the seeing of individual images. The basis functions of this kernel are taken to be the products of polynomials and Gaussian distributions.
	
The steps ISIS performs are as follows: first, all images are transformed to a common template coordinate grid using a polynomial astrometric transformation. Next, several of the images with the best seeing are transformed to the same seeing and stacked to form a reference image. A best-fit kernel is then computed for each image, which is convolved with the reference to match the reference image to the seeing of each image. Each image is then subtracted from the convolved reference image. The subtracted images are stacked to form a median image; bright spots in this image indicate either variable stars or saturated stars, for which the PSF matching cannot be done properly, and which therefore generate residuals. PSF-fitting photometry is performed on each subtracted image to generate a light curve for variable objects.

Using a trimmed version of the image (151 x 301 pixels) with ISIS avoided undesirable effects of the many saturated foreground giant stars in the full (1024 x 2304 pixels) image. Since these trimmed images were a manageable size, they were processed in a single section (\ttfamily{sub\_x = sub\_y = 1}\normalfont).

 Because the coordinates of the two stars of interest, A and B, were already known, ISIS parameters which optimized the variation signal for these sources in the median stacked subtracted image (\ttfamily abs.fits\normalfont) and also minimized contamination by PSF residuals of nearby saturated stars (see section 4) were chosen. Each image was transformed to a common coordinate grid using two-dimensional polynomial astrometric re-mapping of degree 1 (\ttfamily{DEGREE = 1}\normalfont).
 
 In the g' filter, a reference image for subtraction was created using the four images with the best seeing, and in the r' filter, eleven best-seeing images were used. Images were processed using 10 stamps in each direction, each of radius 15 pixels (\ttfamily{nstamps\_x = nstamps\_y = 10, half\_stamp\_size=15}\normalfont), a radius for the convolution kernel of 9 pixels (\ttfamily{half\_mesh\_size = 9}\normalfont), third degree variations in the background level (\ttfamily{deg\_bg=3}\normalfont) and a second degree spatial variation of the kernel (\ttfamily{deg\_spatial=2}\normalfont).The saturation level was left at 1 000 000, as lowering it seemed to contaminate our lightcurves more severely. After image subtraction, median stacking, and variable detection, the ISIS-derived coordinates (in \ttfamily{phot.data}\normalfont) for the centres of A and B were adjusted manually so as to be more aligned with the sources prior to photometry. Photometry was then performed on the subtracted images using a fitting radius of 6 pixels (\ttfamily radphot = 6.0\normalfont). An exploration of the results of varying other photometry parameters led us to conclude that the default values supplied by ISIS were sufficient for our purposes.
 
\section{Light Curve Extraction \& Calibration}\label{s:lightcurve}
\begin{figure}
\figurenum{1}
\includegraphics[scale=0.45]{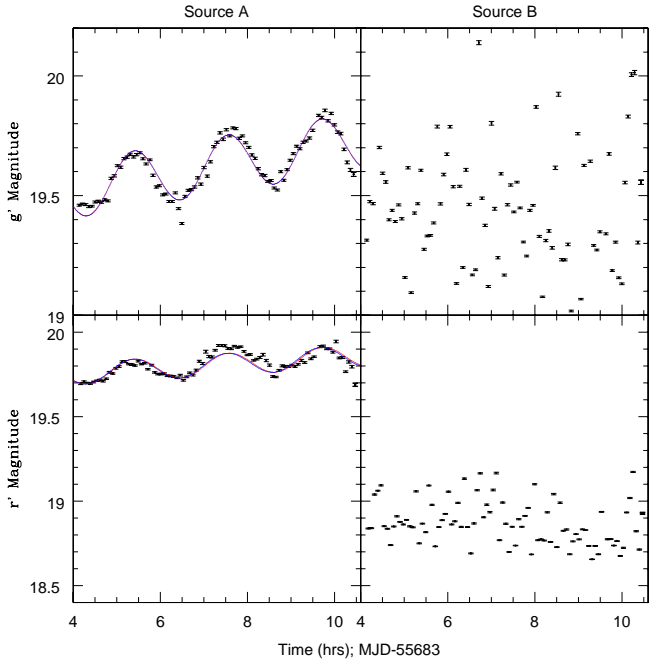}

\caption{{\it Top left:} ISIS g' light curve for A, with two fits which are combinations of a sinusoid, a linear function, and a constant offset. Fitting parameters were computed by QDP for the blue line shown. The (similar) red line is identical save for its period, which is the output from a Lomb-Scargle test. A secular trend of decreasing brightness is also apparent. {\it Bottom left:} The same as above, but for A in the r' filter. The variation is less obvious, but as discussed in the text, the two periods represented by the red and blue lines agree quite closely to those apparent in the g' filter. {\it Top right:} B's g' ISIS light curve. No periodic variation is evident. {\it Bottom right:} B's r' ISIS light curve. Again, no periodic variation is apparent. Errors shown here represent \emph{relative} errors computed by ISIS only; they do not reflect errors associated with the conversion into magnitudes.} 
\label{lightcurve_comparison}

\end{figure}

The lightcurves from the ISIS photometry for A and B appear in Figure \ref{lightcurve_comparison}. Calibration was performed in two steps, according to the method employed by \citet{Mochejska01}; first, the magnitudes of A and B in a template image were computed using the DAOphot package ALLSTAR \citep{Stetson87}. An aperture correction was applied to these magnitudes by comparing the flux admitted through apertures equal to the PSF and 2 pixels larger than the PSF for stars in the vicinity of A and B (central region of the image in Figure \ref{starfield}). 
	
	The template image in both filters was chosen to be one with superior seeing. The template magnitudes, $m_{\rm tpl}$ , for each source were converted into template image counts, $c_{\rm tpl}$, using the ALLSTAR zero point of 25.0 magnitudes and the exposure time of 75.0 s using the relation:
\begin{equation}
m_{\rm tpl} = 25.0 - 2.5\, \log(\frac{ c_{\rm tpl} }{75.0})
\label{eq:mags}
\end{equation}
Once the counts for A and B in the template image were obtained, it was possible to determine their counts in the stacked reference image used for subtraction, $c_{\rm ref}$, by adding the counts on the template image and the delta count value ISIS reported for the subtracted template image,  $\Delta c_{\rm tpl}$: c$_{\rm ref} = c_{\rm tpl} + \Delta c_{\rm tpl}$. To convert the light curve point by point into magnitudes, the count values for A and B were computed for each image by subtracting the image's ISIS delta count values from the reference image count values:
\begin{equation}
c_i = c_{\rm ref} - \Delta c_i
\end{equation}
and inserting the results into equation \ref{eq:mags}. 
Errors in magnitude were computed from the ISIS flux errors using the relation
\begin{equation}
\delta m = 2.5 \, \log(1 \pm \frac{\delta counts}{counts})
\end{equation}
where the upper error bound corresponds to the addition, and the lower, to the subtraction.

The second part of the calibration involved attempting to compute the `true' apparent magnitudes using comparison with photometric standards. Two standard field images, of E8-a and 160100-600000, taken with the same CCD chip on the morning of May 2, 2011, were downloaded from the Gemini archive. The Southern u'g'r'i'z' standard star catalog\footnote{http://www-star.fnal.gov/Southern\_ugriz/www/Fieldindex.html} from \citet{Smith07} was used. A total of 16 standard stars in g' and 20 in r' from across both fields were compared to derive calibration factors of 3.12$\pm$0.08 mags in g' and 3.25$\pm$0.04 mags in r' after weighted averages were taken. It should be noted that the factors derived for each field separately varied on the order of 0.1 magnitude for both filters.  

\section{Light Curves \& Contamination Effects} \label{s:contamination}
\begin{figure}
\figurenum{2}
\includegraphics[scale=0.45]{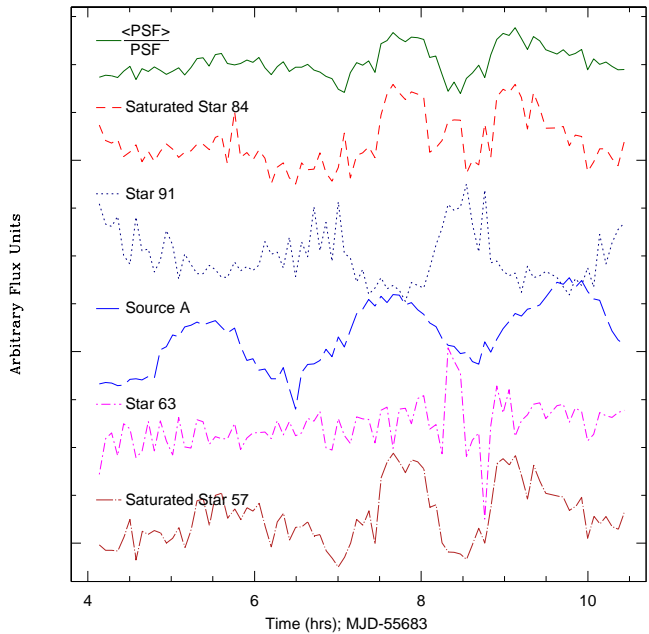}

\caption{Lightcurves for several stars in the g' filter, rescaled and shifted. To emphasize curve resemblances, the \emph{inverse} of the PSF curve is plotted at top. Star 57, a typical example of a saturated star, has a curve which closely mirrors the inverted PSF. Unsaturated stars in the vicinity of saturated stars can also be affected, as seen in Star 37's light curve and Star 91's light curve, which follows the PSF closely. Some unsaturated stars near bright stars do not show this effect, such as Star 63 and, to some extent, Source A.} 
\label{norm_comp}
\end{figure}

While Figure~\ref{lightcurve_comparison} shows no periodic fluctuation in B's light curve, it indicates a clear periodic variation for A in both filters. When A's light curve is compared to the ISIS light curves for other stars, however, it becomes apparent that a possible contamination exists. In Figure \ref{norm_comp}, the g' light curves for several stars, including A, and the inversion of the time variation of the PSF as computed by the IRAF task PSFMEASURE are shown. The PSF has been inverted here for visual aid in identifying curve similarities. Most of the saturated stars in the image displayed light curves that closely resembled the inverse PSF, much like the curve of Star 57 in the figure. This is understandable, as the PSF fitting routine in ISIS cannot properly handle saturated stars, and thus significant ringlike residuals - due to incorrect PSF weighting - appear in the subtracted images. When the PSF is \emph{large}, the computed flux for saturated stars is too \emph{small}.

Light curves of several \emph{un}saturated stars in the vicinity of saturated stars also appear contaminated by the PSF residuals of their neighbours; see Stars 37 and 91 in Figure \ref{norm_comp}. \citet{Hartman04} noted a similar spurious variability introduced in the light curves of stars near saturated sources. Not all unsaturated stars near saturated stars in our images exhibit this effect, however; the curve for Star 63 (see Figure \ref{norm_comp}), which is adjacent to saturated star 57 (see Figure \ref{starfield}), displays little similarity to the PSF curve. Looking at the curve for A, it seems plausible that while the PSF has impacted the result to some degree, a true periodic signal exists. Marked dissimilarities between the locations of the peaks in the PSF curve and source A's curve corroborate this assertion. Results in the r' filter are similar. 

\begin{figure}

\figurenum{3}
\includegraphics[scale=0.45]{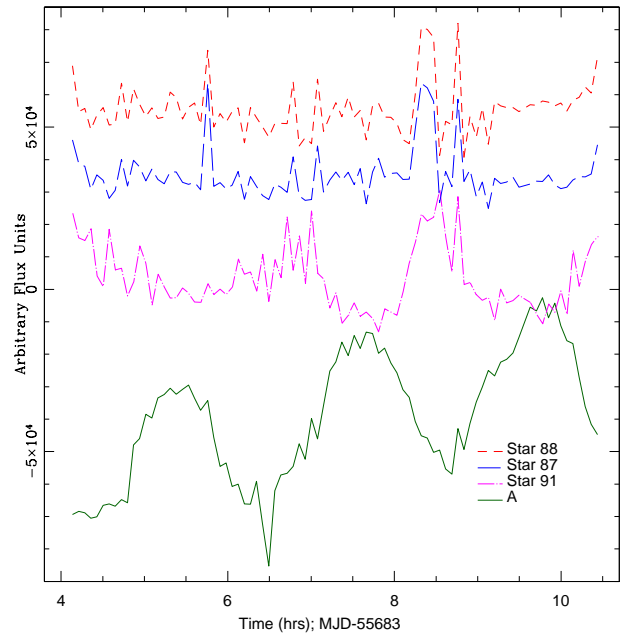}

\caption{Lightcurves for several stars in the g' filter shifted, but not rescaled. Stars 87, 88, and A appear similar in brightness in Figure \ref{starfield}, yet the amplitude of the variations in A's lightcurve are much more significant. Star 91 is much brighter than A, as seen in Figure \ref{starfield}, and yet exhibits smaller amplitude variations, as well. The scatter in the curves for stars 87, 88, and 91 is also much more significant than that for A; this and the amplitude contrast evident above lend confidence to the interpretation of A's variation as a real signal.} 
\label{faint_norm_comp}
\end{figure}

\begin{figure}

\figurenum{4}

\includegraphics[scale=0.45]{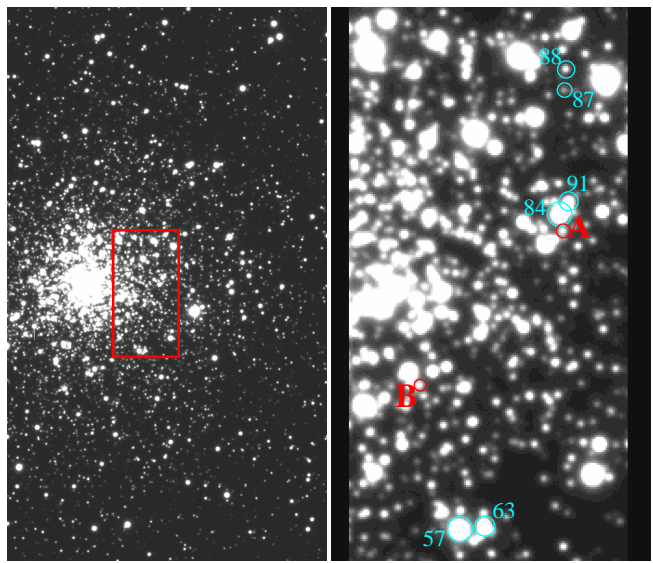}

\caption{{\it Left:} Portion of the image from the central CCD chip of \textit{Gemini's} GMOS-S camera. The rectangle marks the region at right. {\it Right:} Closer view of the region of interest.  Stars whose light curves appear in Figures \ref{norm_comp} and \ref{lomb} are labeled, as well as A and B. The bright star between A and Star 91 is saturated, as are the stars to the left of Star 37. North is toward the top of the image.} 
\label{starfield}
\end{figure}

It is evident from Figure \ref{starfield} that the comparison stars in Figure \ref{norm_comp} are all brighter than A. It is desirable to compare A to a faint star in the vicinity of a saturated star; however, ISIS photometry can be performed only on \textit{variable} stars, and few of the faint stars in the image were detected as variable by ISIS. Two faint stars that \emph{were} picked up by ISIS on the subtracted images, Stars 87 and 88, appear in Figure \ref{faint_norm_comp}. While these are farther from saturated stars than A is, their smaller amplitude variations (compared to A) suggests that A's signal is not spurious. Indeed, Star 91's lightcurve (also shown in Figure \ref{faint_norm_comp}) is also of smaller amplitude than A's, despite its proximity to the saturated Star 84. Finally, all of the comparison lightcurves in Figures \ref{norm_comp} and \ref{faint_norm_comp} possess greater scatter than A's relatively smooth curve. We therefore concluded that the sinusoidal variation in A is a reflection of the binary orbit - possibly distorted somewhat by the saturated stars' PSF residuals - and proceeded to compute a candidate period for the motion. Figure \ref{starfield} presents the star field of interest, along with an enlarged image displaying stars used for comparison in Figure \ref{norm_comp}.

\section{Analysis of Period}\label{s:analysis}
\begin{figure}
\figurenum{5}
\includegraphics[scale=0.48]{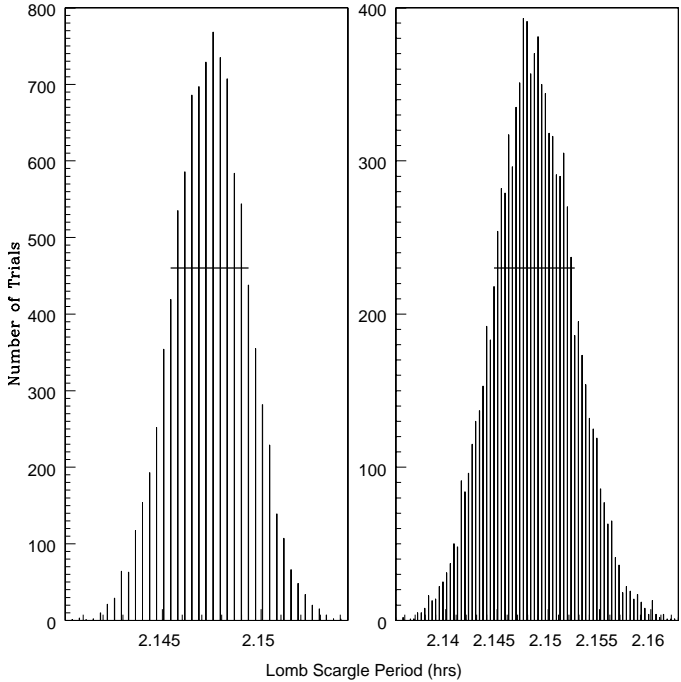}
\caption{{\it Left:} Distribution of periods of 10$^4$ resampling trials for Lomb Scargle period of A using g' filter data. Errors for each data point were multiplied by a factor randomly drawn from a normal distribution about 0 with a standard deviation of 1.0, then added to each data point. A Lomb Scargle test, using an oversampling parameter of 2048, was performed on each resulting data set. The horizontal line shows an approximate 1 $\sigma$ error designation, such that $\sim$15.8\% of the trials lie on either side of the bounds.
{\it Right:} Same as detailed above, but for the r' filter data. } 
\label{montecarlo}

\end{figure}

To determine the period of A's variation, a Lomb-Scargle Periodogram \citep{Lomb76, Scargle82} was calculated for data in each filter. For the g' data, a clear peak appeared at a period of 2.148 $\pm$ 0.002 hours with a false alarm probability (FAP), or probability that the data was drawn randomly from a Gaussian distribution, of  $8.2\times 10^{-12}$. To determine the error in the period, a Monte Carlo method was employed. First, the errors in each data point were multiplied by a factor randomly drawn from a normal distribution about 0 with a standard deviation of 1.0.
The result of this multiplication was then added to each data point, and the Lomb Scargle test was applied to the new data set. This process was repeated 10$^4$ times using an oversampling parameter of 2048, sufficiently high to adequately sample the frequency space. The results for the peak period for each of the 10$^4$ trials were plotted and fit to a pseudo-Gaussian distribution; the values  on both sides of the central peak which were equivalent to the 1$\sigma$ mark in terms of outlying percentage of data ($\sim$15.8\% lying outside these limits on either side) were taken to represent the upper and lower error limits. Figure \ref{montecarlo} displays the results of our Monte Carlo trials.

The above analysis was repeated for the r' filter data to yield a period of 2.149$\pm$0.004 hours, with an FAP of $2.7 \times 10^{-9}$. 

\begin{figure}
\figurenum{6}
\includegraphics[scale=0.48]{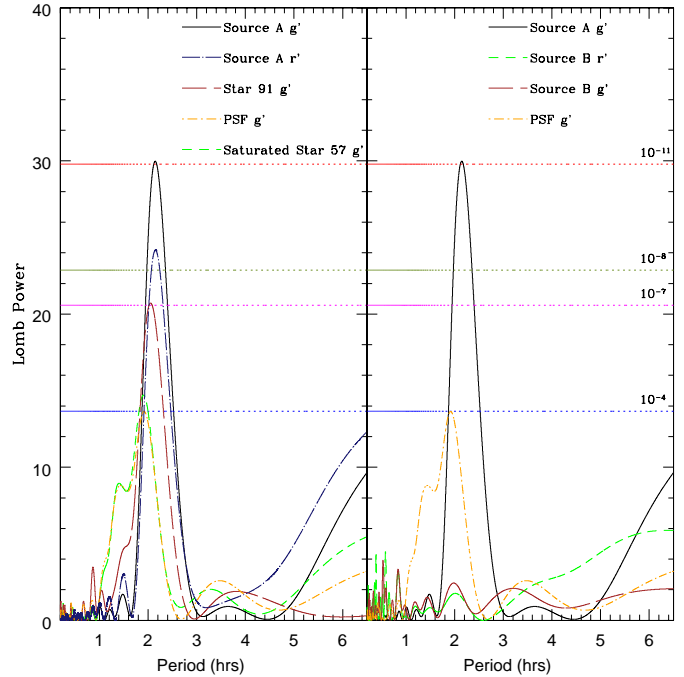}
\caption{{\it Left:} Lomb-Scargle periodogram displaying the significance of the periodicities in the light curves of several stars. The highest peaks belong to A in the g' and r' filters, respectively. 
Also shown is Star 91, an unsaturated star in the vicinity of A and saturated star 84; periodicity in its light curve suggests a contaminating effect of PSF residuals near saturated stars, corroborated by the shape of its lightcurve in Figure \ref{norm_comp}.
A periodicity with FAP $\sim$ 10$^{-4}$ appears in the PSF variation, and the curves of saturated stars, of which star 57 is a typical example, display the same peak. The four dotted horizontal lines mark confidence levels; the FAP for each is indicated. 
{\it Right:} Lomb-Scargle periodograms for B in both filters, accompanied by periodograms for the PSF and A for reference.  No significant periodicity is seen in B's optical light curves. } 
\label{lomb}

\end{figure}

The Lomb-Scargle periodograms for several stars in the field are shown in Figure \ref{lomb}. The two highest peaks correspond to A in the g' and r' filters, respectively.  Star 91, itself unsaturated, is located above the saturated star adjacent to A (see Figure \ref{starfield}), and thus likely suffers from the same PSF contamination that A experienced, leading to the peak in its periodogram somewhere between that of A and the PSF. Lack of complete coincidence with the PSF period peak could be attributed to the photometry of star 91 picking up the periodic signal from A, though the distance between the stars presents an obstacle to this interpretation. In any case, the lightcurves in Figure \ref{norm_comp} present convincing evidence that any periodicity detected in star 91 is spurious, and an effect of contamination. Peaks for other stars we examined were less significant than those shown. For visual aid, several FAP levels are plotted alongside the periodograms in Figure \ref{lomb}. To determine the Lomb power these FAPs correspond to for plotting purposes, the relation
\begin{equation}
P_{\rm FAP}= 1 - (1 - e^{-z})^{M}
\label{eq:fap}
\end{equation}
was used, where P$_{\rm FAP}$ is the false alarm probability, z is the Lomb power, and M is equivalent to the number of data points for our purposes \citep{Press92}.

No significant peaks appeared when a Lomb-Scargle test was performed on B's light curves, as shown by the right panel in Figure \ref{lomb}.

A separate method of fitting the light curves was pursued to obtain an independent estimate of the periodicity. The plotting program QDP\footnote{http://heasarc.gsfc.nasa.gov/docs/software/ftools/others/qdp/qdp.html} was used to fit A's light curves to the functional form
\begin{equation}
A \sin(\frac{2\pi}{B}(x+C)) + Dx + E.
\label{func_form}
\end{equation}
Due to an initially large reduced $\chi^2$ value obtained upon fitting, the errors were increased until the $\chi^2$ was comparable to the degrees of freedom. Fitting the data set with seven times the initial errors yielded parameters A =  0.119$\pm$0.008; B = 2.15$\pm$0.02; C = 7.01$\pm$0.02; D = 0.031$\pm$0.003; and E = 19.40$\pm$0.02. This fit is plotted in Figure \ref{lightcurve_comparison} in blue. The same fit with the period changed to the result of the Lomb-Scargle test for comparison is shown in red. 

An identical fitting procedure was applied to the r' filter data, with the results A = 0.065$\pm$0.009; B = 2.16$_{-0.03}^{+0.04}$; C=0.5$\pm$0.2; D = 0.016$\pm$0.004; and E = 19.69$\pm$0.03. Errors here are reported at the 1$\sigma$ level. Again, the fit (blue), along with an altered version containing the Lomb-Scargle period (red), appears in Figure \ref{lightcurve_comparison}.

We are confident in our estimate of the candidate period due to the agreement of the Lomb-Scargle period estimations in both filters with one another and the agreement of the QDP fit results with the Lomb-Scargle estimates within 1$\sigma$.

\begin{figure}
\figurenum{7}

\includegraphics[scale=0.48]{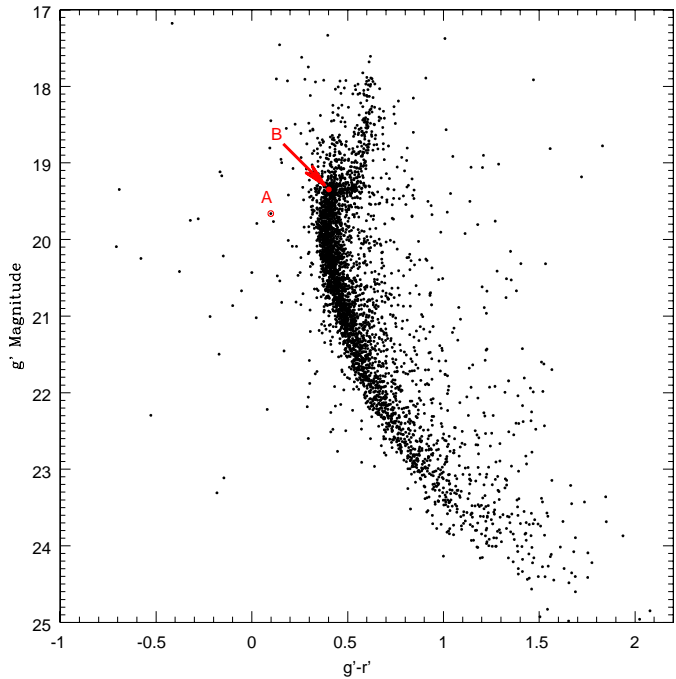}

\caption{Color-Magnitude Diagram constructed using consecutive images in r' and g' with good seeing. All stars within 66.5\arcsec of the cluster centre were included in the image, and the breadth of the main sequence indicates the typical errors. Due to large $ \chi ^2$ residuals given by ALLSTAR for A and B, their magnitudes should be regarded as suffering larger errors.}
\label{cmd}
\end{figure}

\section{Color Magnitude Diagram}\label{s:cmd}
The DAOphot package ALLSTAR was used to perform PSF fitting photometry on the best-seeing images.  Color-magnitude diagrams were constructed using images taken consecutively in g' and r', separated 
by $\sim$four minutes. Due to severe crowding by saturated stars (see Figure \ref{starfield}), we could obtain usable photometry for A and B in only a few images. Figure \ref{cmd} shows our colour-magnitude diagram with the tightest main sequence that shows A and B. A caveat is that the reduced $\chi^2$ values for A and B, indicative of goodness of PSF fit in DAOphot, were markedly higher (7.0, 5.1, 7.1 and 3.8 for A and B in r' and g', respectively) than for most main sequence stars (typically 1.0), indicating that the photometry for A and B is less reliable than that for most main sequence stars.
 
The position of the main sequence turn-off is comparable to that in \citet{Heinke01}. B again lies on the main sequence, while A is bluer than it, similar to results in those authors' \textit{HST} color-magnitude diagram.\section{Discussion}\label{s:discuss}

The possible contamination of A's source lightcurve by saturated stars forces us to exercise caution in our attribution of a period.  However, several arguments point to the 2.15 hour period being real. The Lomb power peak for the PSF modulations and other saturated stars was $\sim$1.9 hours, significantly different from A's Lomb peak. Figure \ref{norm_comp} shows that the shape of the lightcurve due to the PSF modulation is significantly different from that of A, and Figure \ref{faint_norm_comp} displays A's larger amplitude and smoother variation with respect to stars similarly situated. A sinusoidal shape is a very good description of A's lightcurve, as seen in Figure \ref{lightcurve_comparison}.  Finally, the 2.15 hour period is seen clearly in both $g'$ and $r'$, and is consistent with the longer of the two candidate periods for A in \citet{Heinke01}. \citet{Heinke01}'s paper represents an independent dataset confirming our own observations; furthermore, since their data was taken with \textit{HST}, for which neither seeing nor saturated stars were an issue, the agreement lends weight to our interpretation of the 2.15 hour period as a real signal. A repeat observation, using a shorter frame time to prevent nearby stars from saturating, could give final confirmation of this period. 


A 2.15 hour period and A's persistent X-ray luminosity of $\sim10^{36}$ ergs/s for 20 years suggest a system similar to GS 1821-26 \citep{Homer98,Meshch10} and a donor star of order 0.2 \Msun \citep[cf.][]{Deloye08b}.  

A sinusoidal modulation of $\sim0.1$ magnitude could be due to heating of one face of the donor, or ellipsoidal modulations due to the varying donor aspect, as often seen for cataclysmic variables \citep{Edmonds03b}, which would imply a 4.3 hour period.  We rule out ellipsoidal variations on two counts, however: firstly, the larger amplitude of the $g'$ vs. $r'$ modulations of A imply that the surface with changing visibility is quite hot, like the face of the donor heated by the disk. Further, a 0.2 \Msun star should be located near the bottom of the main sequence. A's position (see Figure \ref{cmd}) suggests that there must be substantial contributions to the luminosity from the bluer accretion disk and the heated face of the secondary. The \emph{relative} variation of flux, and therefore the magnitude variation, is much smaller than it would be were no disk visible. \cite{Wang04} outline a case with a similar variation amplitude corresponding to sinusoidal variations. The small amplitude of A's variations may be explained by emission of the disk overwhelming the sinusoidally modulated visibility of the hot side of the donor star. We therefore think the identification of 2.15 hours as the binary orbital period is solid.



A's position in the $g'$, $g'-r'$ color magnitude diagram is slightly lower than the main sequence turnoff, in contrast with its position slightly above the turnoff in a $V$, $V-I$ diagram by \citet{Heinke01}.  We expect that the majority of the light in this system is produced by the hot accretion disk, so the X-ray decay observed in 2011 (see Introduction) may explain the decrease in the optical brightness.



B's lightcurve (Figure \ref{lightcurve_comparison}) appears completely chaotic. While this suggests that ISIS may not be producing reliable results, this flickering is more likely to be intrinsic to the system. The flickering is larger in $g'$ ($\sim$1 mag) than $r'$ ($\sim$0.5 mag), which is expected if it relates to X-ray reprocessing, but not for saturation effects (as nearby giants are more saturated in $r'$). The flickering is not unlike the variations seen by \citet{Deutsch00} in B's HST lightcurve, both in amplitude ($\sim$1 mag in $V$ and $I$) and timescales (5-20 minutes).  Finally, B's \Chandra\ X-ray lightcurve shows order-of-magnitude flaring with typical timescales of 5 minutes \citep{Coomber11,Stacey11b}.  Considering these arguments, we think that the ISIS photometry for B may be accurately reflecting the optical variation of the system, making it an intriguing target for future simultaneous X-ray/optical studies.  If B's photometry presented here is reliable, it strongly indicates that the 43 minute candidate orbital period of \citet{Deutsch00} is spurious.

The optical color of B remains a significant puzzle.  Since B's flickering occurs on the same timescales as our exposure lengths, it's difficult to accurately measure its color; however, the g' and r' measurements suggest a color as red as, or redder than, the main sequence.  The flaring requires some component of the system--the disk, the companion, or both--to be strongly heated by X-rays, which should make it blue.  Thus, B's color suggests a donor redder than the main sequence, perhaps a subgiant or "red straggler" star with an unusual evolutionary history \citep{Albrow01,Ferraro01b,Mathieu03}.



\section{Conclusion}\label{s:conclude}

We have identified a clear 2.15 hour sinusoidal modulation in the $g'$ and $r'$ lightcurves from the LMXB A in NGC 6652.  Although contamination of the ISIS lightcurves due to signals from saturated stars is possible, we are confident the sinusoidal signal is robust and likely represents the binary orbital period for reasons outlined above. Further Gemini imaging, with shorter exposure times to avoid saturating nearby stars, could finalize our result.  We note that this period, if confirmed, will be only the second X-ray binary in a globular cluster to have its orbital period measured using ground-based telescopes, after (the much brighter) AC 211 in M15 \citep{Ilovaisky93}.

The low-luminosity LMXB, B, in NGC 6652 shows a highly variable lightcurve with strong $\sim$1 magnitude variations on timescales of 5-20 minutes.  This optical flickering (if real) is probably driven by the strong, rapid X-ray flaring seen from B.

\acknowledgements
We acknowledge support by NSERC and an Alberta Ingenuity New Faculty Award, and useful conversations with C. Deloye.  These results are based on observations obtained at the Gemini South Observatory (Proposal ID GN-2011A-Q-20). The Gemini Observatory is operated by the Association of Universities for Research in Astronomy, Inc., under a cooperative agreement with the NSF on behalf of the Gemini partnership: the National Science Foundation (United States), the Science and Technology Facilities Council (United Kingdom), the National Reseach Council (Canada), CONICYT (Chile), the Australian Research Council (Australia), CNPq (Brazil) and CONICET (Argentina). We thank R. Wijnands for drawing our attention to the decline in the bulge scan lightcurve of NGC 6652.

{\it Facilities:} \facility{Gemini:South (GMOS)}

\bibliography{src_ref_list}
\bibliographystyle{apj}

\end{document}